\documentclass[aps,prl,twocolumn,superscriptaddress,showpacs]{revtex4}

\usepackage{epsfig}
\usepackage{subfigure}
\usepackage{amssymb}
\usepackage{graphicx}
\usepackage{multirow}
\usepackage{array}
\usepackage{amsmath}
\usepackage{natbib}
\newcommand{\mb}{\boldsymbol}
\begin{document}
%\preprint{}

\title{Coherent Propagation of Spin Helices in a Quantum-Well Confined Electron Gas}

\author{Luyi Yang}\affiliation{Department of Physics, University of California,
Berkeley, California 94720, USA.} \affiliation{Materials Science
Division, Lawrence Berkeley National Laboratory, Berkeley,
California 94720, USA.}
\author{J. D. Koralek} \affiliation{Materials Science
Division, Lawrence Berkeley National Laboratory, Berkeley,
California 94720, USA.}
\author{J. Orenstein}
\affiliation{Department of Physics, University of California,
Berkeley, California 94720, USA.} \affiliation{Materials Science
Division, Lawrence Berkeley National Laboratory, Berkeley,
California 94720, USA.}
\author{D. R. Tibbetts}
\author{J. L. Reno}
\author{M. P. Lilly}
\affiliation{Sandia National Laboratories, Albuquerque, New Mexico
87123, USA.}

%\date{\today}

\begin{abstract}
We use phase-resolved transient grating spectroscopy to measure the propagation of spin helices in a high mobility $n$-GaAs/AlGaAs quantum well with an applied in-plane electric field.  At relatively low fields helical modes crossover from overdamped excitations where the spin-precession period exceeds the spin lifetime, to a regime of coherent propagation where several spin-precession periods can be observed. We demonstrate that the envelope of a spin polarization packet reaches a current-driven velocity of 10$^7$ cm s$^{-1}$ in an applied field of 70 V cm$^{-1}$.
\end{abstract}

%\keywords{}

\maketitle
%\end{CJK*}

The properties of materials in which the electron's spin is strongly coupled to its motion are receiving increasing attention in a variety of contexts.  From an applications point of view, spin-orbit (SO) coupling provides a mechanism by which spin polarization lifetime and mobility can be controlled by applied electric fields.  From a basic science perspective, SO coupling introduces many of the phenomena usually associated with time-reversal breaking, such as spin precession \cite{Rashba}, nonuniversal \cite{Sinova,Murakami}, and quantum Hall effects \cite{KM,HgTe}, without the need for large externally applied magnetic fields.

In systems without inversion symmetry, the SO interaction energy is a linear function of the electron wave vector, $\mb{k}$, and the coupling can be viewed as a Zeeman interaction between the spin of the electron and a momentum-dependent effective magnetic field, $\mb{b}(\mb{k})$. SO interactions that are linear in $\mb{k}$ generate a variety of unique nonequilibrium transport properties in both dilute electron gases and degenerate Fermi liquids.  In the single-electron regime, the spin of a ballistic electron precesses coherently about $\mb{b}(\mb{k})$, undergoing a full period of rotation after propagating a length $l_s$ that is inversely proportional to the coupling strength.  The ability to control the strength of the coupling with an externally applied electric field, and consequently $l_s$, is the basis of the proposed spin transistor \cite{DD}.  In the Fermi liquid regime, the spin polarization, $\mb{S}$, and charge density, $n$, are collective properties that are coupled by the SO interaction.  The nonequilibrium dynamics of the system are described by four normal modes for each wave vector, with normal mode coordinates that are admixtures of the four degrees of freedom: $n$ and the three components of $\mb{S}$ \cite{Burkov,Mish,SU2,cubic}.

In the case of two-dimensional SO systems, such as semiconductor quantum wells, oxide interfaces, and thin metallic films, $\mb{b}(\mb{k})$ is confined to the conducting plane.  For isotropic planes or high-symmetry directions, the four normal modes break into two pairs.  One pair describes the coupling of the current to an in-plane component of spin, the other describes the coupling of the remaining in-plane component to the out-of-plane spin, $S_z$. The normal mode solutions to this latter pair of equations are two helical spin polarization waves, with an opposing sense of rotation.  The lifetime of the helix whose sign of rotation matches that of a ballistically propagating electron is enhanced by the SO coupling, while the lifetime of the other helix is reduced. The lifetime-enhanced helix with wavelength equal to $l_s$ becomes a conserved quantity when the two contributions to $\mb{b}(\mb{k})$, known as the Rashba \cite{Rashba} and Dresselhaus  \cite{Dresselhaus} terms, are equal \cite{SL}, reflecting a recovered SU(2) symmetry at this special point in parameter space \cite{SU2}. Spin helices with strongly enhanced lifetimes have been observed by transient grating spectroscopy \cite{Koralek} and subsequently imaged by Kerr microscopy \cite{IBM}.

For a Fermi sea at rest with respect to the lattice, the modes described above are overdamped, that is spin density fluctuations decay exponentially, with a lifetime that depends on the distance in parameter space of the SO Hamiltonian from the SU(2) point \cite{SU2,cubic}.  However, it has been predicted that helical modes become underdamped coherent excitations in a drifting Fermi sea \cite{KB,Yang1}, as would result from an electric field applied in the plane, for example.  In this regime, the spin polarization helix propagates at least one wavelength before it decays and the local spin density acquires an oscillatory component. This form of coherent propagation is essential in order for spins to ultimately play a role in information processing. As the theoretical predictions are based on simplified models that neglect electron-electron and electron-phonon coupling, it is an open question as to whether such modes exist in real, interacting many-body systems.

GaAs quantum wells are ideal model systems in which to detect coherently propagating spin helices, for several reasons.  First, the optical orientation effect in III-V semiconductors enables photoexcitation of nonequilibrium waves of spin density by interfering two orthogonally polarized beams at the sample surface - yielding a so-called transient spin grating \cite{Cameron}.  Second, the strength of the SO coupling in GaAs is such that the micron-scale pitch of helices with enhanced lifetime is well matched to the wavelength of the 1.5 eV photons that are needed to generate the nonequilibrium spin density.  As a result of this matching, the amplitude of the spin polarization can be measured by diffracting a probe beam from the transient grating. We use optical heterodyne detection to measure the phase as well as the amplitude of the diffracted probe \cite{Goodno,Maznev,Nuh1}, as the former is sensitive to the translational motion of the spin helix \cite{Yang2}. Finally, and crucially for the experiments reported here, the combination of low carrier density and high mobility below 70 K enable the generation of large drift velocities with minimal Joule heating.

Phase-resolved transient grating measurements were performed on 9 nm wide $n$-doped GaAs/AlGaAs quantum wells, grown by molecular beam epitaxy on a semi-insulating GaAs (001) substrate (VB0355).  To simplify the spin Hamiltonian, we fabricated symmetric QW structures such that the Rashba interaction is near zero and the Dresselhaus coupling dominates ($\beta_1 = 3.4 \times 10^{-3}$~eV{\AA}~\cite{Yang2}). The two-dimensional electron gas (2DEG) channel was defined by mesa etching, and Ohmic contact was made by annealing NiGeAu to the sample. After patterning, the GaAs substrate was mechanically lapped and chemically etched to allow for optical measurement in transmission geometry. Silicon donors were symmetrically doped in the center of each barrier, yielding a carrier density $1.9\times10^{11}$ cm$^{-2}$ and mobility $5.5\times10^5$ cm$^2$V$^{-1}$s$^{-1}$ at 5 K. Spin density waves with out-of-plane polarization and wave vector, $\mb{q}$, along the [110] crystal axis were photoinjected by 100 fs pulses from a Ti:sapphire laser focused to deliver an intensity of 80 nJ cm$^{-2}$.  Drift motion of the 2DEG parallel to $\mb{q}$ was induced by in-plane electric fields, $\mb{E}$, applied parallel to $\mb{q}$, with variable strength up to 70 V cm$^{-1}$.

Following pulsed photoinjection, the amplitude and position of spin density waves were sensed by mixing the diffracted component of a time-delayed probe beam with a beam of transmitted pulses in a Si photodiode. If the spin wave propagates uniformly with velocity $\mb{v}$, the photodiode output signal $V(t)$ will be proportional to $S_z(\mb{q};t)\cos[\phi(t)+\phi_{pld}]$, where $t$ is the time delay between the arrival of the photoinjection and probe beams, $S_z(\mb{q};t)$ is the amplitude of the wave, $\phi(t)=\mb{q}\cdot \mb{v}t$ is the phase shift caused by translation of the wave along the direction of its wave vector, and $\phi_{pld}=2\pi\lambda/d$ is an adjustable phase proportional to the path length difference between the transmitted and diffracted beams.  For a spin wave that propagates uniformly with exponentially decaying amplitude, the photodiode signal will have the form of a damped oscillation, that is, $V(t)\sim S_z (\mb{q};0)\exp[-\gamma(q)t]\cos(\mb{q}\cdot\mb{v}t+\phi_{pld})$.

\begin{figure}
    \centering
      \includegraphics[width=0.48\textwidth]{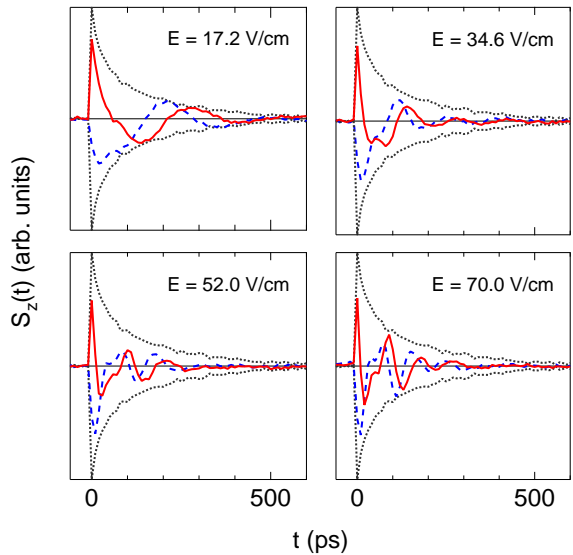}
            \caption{Time evolution of a transient spin grating for various applied electric
            fields at $q=1.07\times10^4$ cm$^{-1}$ and $T=10$ K. The black dotted lines show the decay of the amplitude in zero field (the negative of the amplitude is shown as well as a guide to the eye). The red and blue curves are
            the spin grating with $\phi_{pld}$ set to 0 and $\pi/2$, respectively.}\label{fig1}
\end{figure}

Figure \ref{fig1} shows the time evolution of a $q =1.07\times10^4$ cm$^{-1}$ spin wave, photoinjected into the 2DEG held at 10 K, as recorded with the technique described above.  The dotted lines illustrate the exponential decay observed when the applied electric field is zero.  (Both the amplitude and its negative are shown as a guide to eye in interpreting the signals observed with nonzero field).  The red and blue curves are the signals recorded with nonzero $\mb{E}$ and $\phi_{pld}$ set to 0 and $\pi/2$, respectively. The oscillations that appear with application of $\mb{E}$ clearly demonstrate coherent propagation of spin density waves.  Roughly speaking, each period of the oscillations corresponds to a translation of the transient spin grating by one wavelength.  The drift velocity can be estimated directly from the raw data; for example, at $E=17.2$ V cm$^{-1}$ the period is of order 250 ps, during which time the spin wave shifts by approximately 6 microns, corresponding to a drift velocity of $\sim2\times10^6$ cm s$^{-1}$. Even at this relatively low electric field, the velocity of the current-driven spin texture is quite large, as compared, for example, to driven domain walls in ferromagnets where the typical maximum velocity is $\sim10^4$ cm s$^{-1}$ \cite{Parkin}.

\begin{figure}
    \centering
     \includegraphics[width=0.40\textwidth]{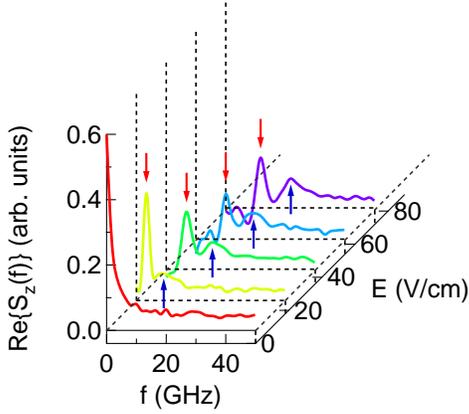}
      \caption{The real part of the Fourier transform of the in-phase component of $S_z(t)$ for various $E$ at 10 K, as a function of frequency, $f$. Two peaks (indicated by the red and blue arrows) are observed for each value of $E$. Each peak frequency is the inverse of the time required for a spin helix to propagate a distance equal to its wavelength.}
      \label{fig2}
\end{figure}

Closer inspection of the curves in Fig. \ref{fig1}, particularly at higher fields, indicates additional structure is present that cannot be described by a single damped sine or cosine function.  To better understand the origin of these features, we Fourier transform the data from time to frequency domain; the real part of the transform of the $\phi_{pld}=0$ (red curves) is plotted in Fig. \ref{fig2}.  As is apparent from the spectra, the structure in the time domain reflects the fact that there are actually two propagating modes at this wave vector, which become more clearly resolved with increasing $\mb{E}$.

\begin{figure*}
    \centering
      \includegraphics[width=0.80\textwidth]{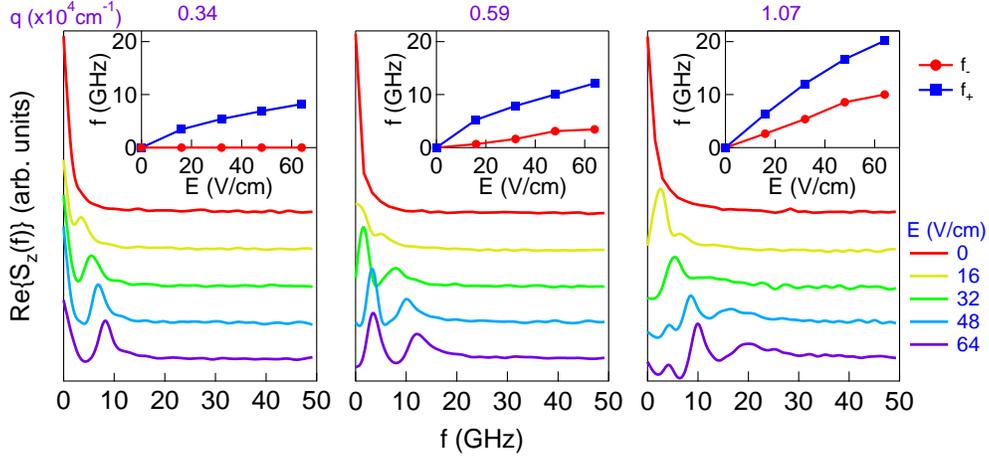}
        \caption{The three panels show the Fourier transform of $S_z(t)$ as a function of $f$ for several values of $E$ at $T=30$ K, for (left to right) $q$=0.34, 0.59, and 1.07$\times$10$^4$ cm$^{-1}$. Inset: the frequencies of the two peaks as a function of $E$. }\label{fig3}
\end{figure*}

Figure \ref{fig3} illustrates how the spectra shown in Fig. \ref{fig2} vary with wave vector.  For each value of $\mb{q}$, two modes are seen, again most clearly resolved at the highest field.  The inset of each panel shows the increase of the frequency of the two peaks with increasing $\mb{E}$. The interpretation of the two collective modes observed with nonzero $\mb{E}$ follows directly from our understanding of the modes with $\mb{E}$ = 0.  The out-of-plane polarized spin wave that is photogenerated at time delay zero is an equal weight superposition of the two oppositely oriented helical normal modes \cite{Weber}.  In zero field, as stated previously, the two photoinduced helices decay exponentially, with different lifetimes.  When photoinjected into a drifting Fermi sea our results show that both helices propagate coherently, but at different velocities, yielding the two resonant frequencies for each wave vector.

%\begin{figure*}
%    \centering
%    \subfigure[]{
%      \label{fig4a}
%      \includegraphics[width=0.33\textwidth]{fig4a1.eps}}
%    \subfigure[]{
%      \label{fig4b}
%      \includegraphics[width=0.33\textwidth]{fig4b1.eps}}
%      \caption{(Color online) (a) The dispersion relations of the resonant frequencies associated with the two helical modes, $f_{\pm}$, at $E=64$ Vcm$^{-1}$ and $T=30$ K. The solid lines are a linear fit. (b) The group velocity $v_g$ of the spin packet as a
%        function of the applied $E$ field at $q=1.07\times10^4$ cm$^{-1}$ and $T=10$ K. The dashed line is the extrapolation of the linear response regime at low $E$, i.e. $v_g=\mu_sE$.}\label{fig4}
%\end{figure*}

\begin{figure}
    \centering
    \subfigure[]{
      \label{fig4a}
      \includegraphics[width=0.23\textwidth]{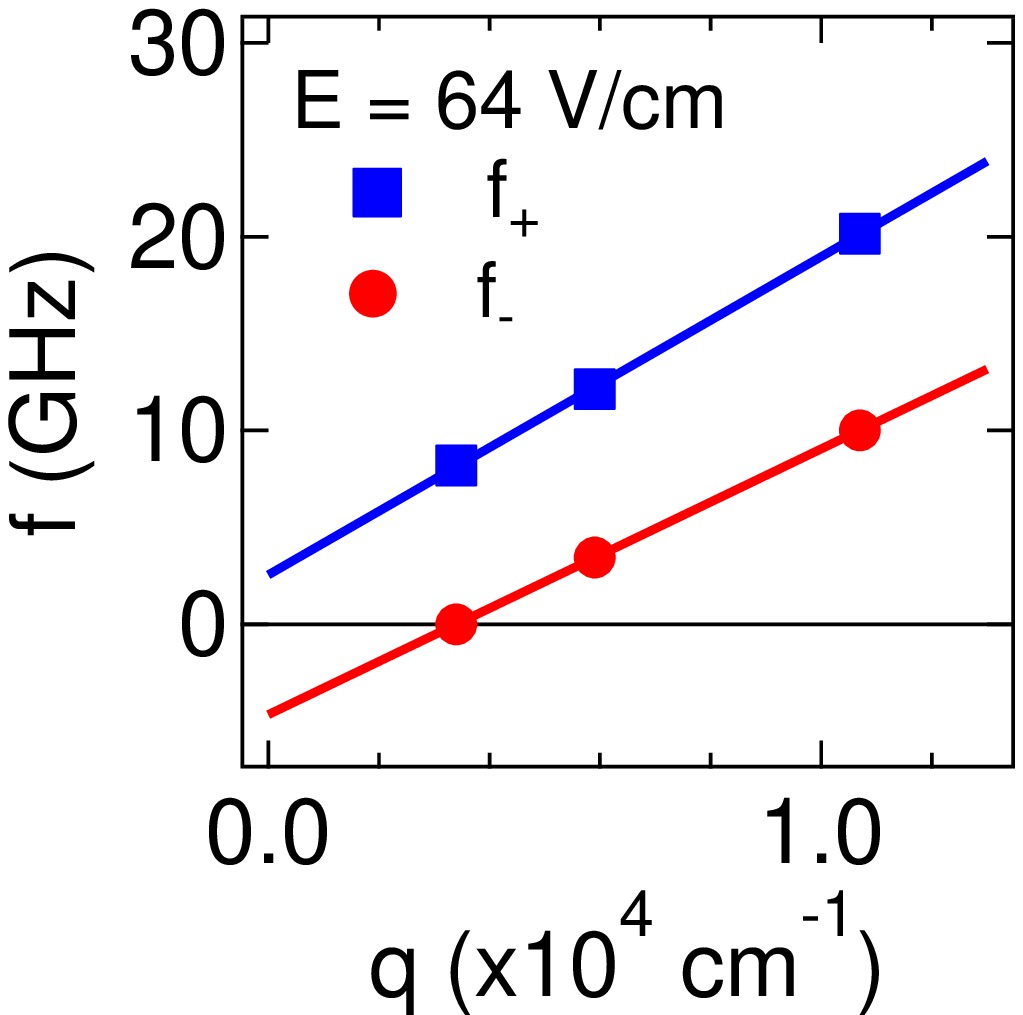}}
    \subfigure[]{
      \label{fig4b}
      \includegraphics[width=0.23\textwidth]{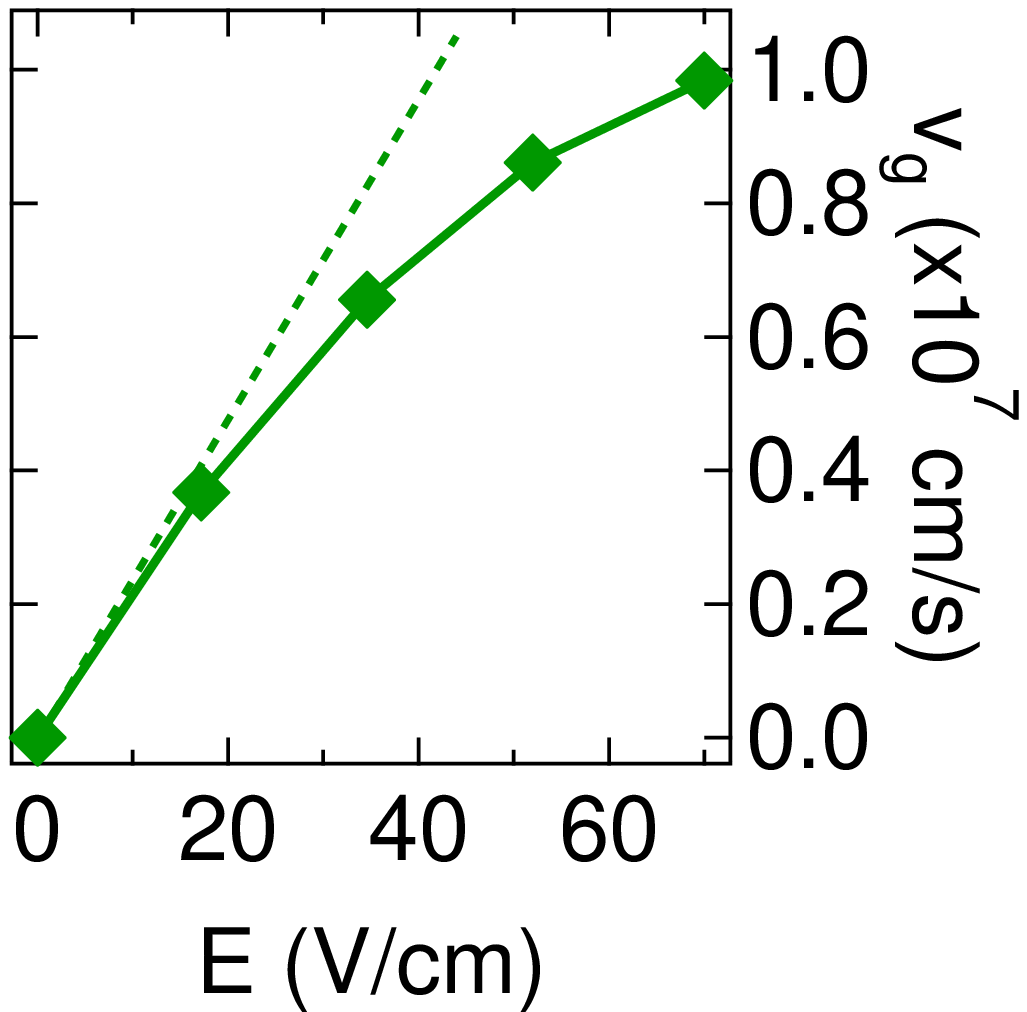}}
      \caption{(a) The dispersion relations of the resonant frequencies associated with the two helical modes, $f_{\pm}$, at $E=64$ V cm$^{-1}$ and $T=30$ K. The solid lines are a linear fit. (b) The group velocity $v_g$ of the spin packet as a
        function of the applied $E$ field at $q=1.07\times10^4$ cm$^{-1}$ and $T=10$ K. The dashed line is the extrapolation of the linear response regime at low $E$, i.e., $v_g=\mu_sE$.}\label{fig4}
\end{figure}

In Fig. \ref{fig4a}, we plot the frequencies, $f_\pm$, of the two modes as a function of wave vector for $E=64$ V cm$^{-1}$.  Both helices disperse linearly and with very nearly the same slope, in qualitative agreement with theoretical predictions for the modes of a drifting 2DEG in the presence of SO coupling \cite{KB,Yang1}. In particular, Kleinert and Bryksin \cite{KB} (KB) obtained a dispersion relation of the form, $2\pi f_\pm(q) =i\gamma_\pm (q)+v_d (q\pm q_0)$, where $v_d$ is the drift velocity, $\gamma_\pm$ are the helix decay rates, and $q_0$ is the wave vector at which the lifetime of the SO stabilized helix is maximal. The KB dispersion relation is somewhat unusual, as it predicts that the longer-lived, (-), helix is stationary when $q=q_0$ and actually propagates in the direction opposite to the 2DEG for $q<q_0$.  However, the propagation of the helical pattern is not equivalent to the velocity of the envelope of a packet of spin polarization.  For example, a helical pattern of noninteracting localized spins in an applied magnetic field will appear to propagate with $v=\omega_Z/q$, where $\hbar\omega_Z$ is the Zeeman energy,  whereas an envelope of spin polarization would be immobile.

Applying the usual analysis of wave packet motion to the KB dispersion relation shows that it is the quantity $\partial\omega/\partial q$ (where $\omega=2\pi f)$, rather than the frequencies of the modes themselves, that determines the group velocity, $v_g$, at which a spin polarized wave packet will propagate.   In Fig. \ref{fig4b}, we plot $v_g$ as a function of $E$, as determined from the average value of $\partial\omega/\partial q$ for the two modes.  The group velocity increases linearly at first and then begins to saturate with further increase  of $E$, nevertheless reaching $\sim$ 10$^7$ cm s$^{-1}$ at 70 V cm$^{-1}$, which is approximately the Fermi velocity.  In the linear regime at low $E$, the spin mobility $\mu_s\equiv v_g/E\approx2.5\times10^5$cm$^2$V$^{-1}$s$^{-1}$, which is roughly half of the electron mobility as determined from dc transport methods. This difference is consistent with the observation \cite{Yang2} that $v_g/v_d$ depends on the photoinjected electron density, $\Delta n$, approaching unity only in the limit that $\Delta n/n\rightarrow 0$.

Our results demonstrate that overdamped modes of a spin-orbit coupled 2DEG in GaAs crossover to coherently propagating helical waves when the spin-precession period becomes smaller than the spin-relaxation time, which for our sample occurs in the presence of modest electric fields  $\sim$0.2 V applied across a 200 micron channel.  In the Dresselhaus-coupled system studied here, electron spins precess $\sim$3 full revolutions within the 200 ps polarization lifetime, during which time a wave packet of spin polarization will propagate $\sim$25 microns.  These results suggest that controlling the coherence length of spin transport, with large dynamic range, can be achieved by adding Rashba SO coupling via out-of-plane electric fields. For example, a $\times$50 variation of helix lifetime as a function of electric fields applied by asymmetric doping has been demonstrated \cite{Koralek} and, in theory, this dynamic range can be exceeded with fields applied by an external gate electrode.  Finally, the phenomena that we have observed in semiconductor quantum wells should arise in all inversion breaking SO systems, of particular interest are those in which stronger coupling implies nanoscale precession lengths and precession rates in the terahertz regime.

\begin{acknowledgments}
All the optical and some of the electrical measurements were carried
out at Lawrence Berkeley National Laboratory and were supported by
the Director, Office of Science, Office of Basic Energy Sciences,
Materials Sciences and Engineering Division, of the U.S. Department
of Energy under Contract No. DE-AC02-05CH11231. Sample growth and
processing and some of the transport measurements were performed at
the Center for Integrated Nanotechnologies, a U.S. Department of
Energy, Office of Basic Energy Sciences user facility at Sandia
National Laboratories (Contract No. DE-AC04-94AL85000).
\end{acknowledgments}

\end{document}